\pdfoutput=1
\documentclass[twocolumn,preprintnumbers,superscriptaddress,aps]{revtex4-1}

\usepackage{amsthm,amsmath,amssymb,physics,mathtools}
\usepackage{bbm}

\usepackage{array,esint}
\def\CMD#1{%
   $ \csname#1\endcsname \displaystyle\csname#1\endcsname $ & \texttt{\textbackslash#1} &}

\usepackage{hyperref}
\hypersetup{
    colorlinks = true,       % false: boxed links; true: colored links
    linkcolor = red,          % color of internal links (change box color with linkbordercolor)
    citecolor = magenta,        % color of links to bibliography
    filecolor=red,      % color of file links
    urlcolor= cyan           % color of external links
}

\usepackage{graphicx,tikz}
\graphicspath{ {TBA-PRL/} }
\usepackage{xcolor}

\usepackage{dsfont}

\newcommand{\be}{\begin{equation}}
\newcommand{\ee}{\end{equation}}
\newcommand{\beq}{\begin{equation}}
\newcommand{\eeq}{\end{equation}}
\newcommand{\bea}{\begin{equation} \begin{aligned}}
\newcommand{\eea}{\end{aligned} \end{equation}}

\def\C{\mathcal{C}}
\def\D{\mathcal{D}}
\def\E{\mathcal{E}}

\def\L{\mathcal{L}}
\def\N{\mathcal{N}}
\def\Op{\mathcal{O}}

\def\bQ{\bold{Q}}

\def\bT{\bold{T}}
\def\bt{\bold{t}}

\def\WR{\mathbb{W}^{R}}
\def\WL{\mathbb{W}^{L}}
\def\WB{\mathbb{W}^{B}}

\def\bW{\mathbb{W}}
\def\Y{\mathcal{Y}}

\def\bz{\bold{z}}

\def\su{\mathtt{su}}

\def\pA{\hat{\mathtt{p}}}
\def\pS{\tilde{\mathtt{p}}}

\def\p{p^{\textrm{mir}}}
\def\m{\mu^{\textrm{mir}}}

\def\Xint#1{\mathchoice
   {\XXint\displaystyle\textstyle{#1}}%
   {\XXint\textstyle\scriptstyle{#1}}%
   {\XXint\scriptstyle\scriptscriptstyle{#1}}%
   {\XXint\scriptscriptstyle\scriptscriptstyle{#1}}%
   \!\int}
\def\XXint#1#2#3{{\setbox0=\hbox{$#1{#2#3}{\int}$}
     \vcenter{\hbox{$#2#3$}}\kern-.5\wd0}}

\def\dashint{\Xint-}

\DeclareMathOperator*{\SumInt}{%
\mathchoice%
  {\ooalign{$\displaystyle\sum$\cr\hidewidth$\displaystyle\int$\hidewidth\cr}}
  {\ooalign{\raisebox{.14\height}{\scalebox{.7}{$\textstyle\sum$}}\cr\hidewidth$\textstyle\int$\hidewidth\cr}}
  {\ooalign{\raisebox{.2\height}{\scalebox{.6}{$\scriptstyle\sum$}}\cr$\scriptstyle\int$\cr}}
  {\ooalign{\raisebox{.2\height}{\scalebox{.6}{$\scriptstyle\sum$}}\cr$\scriptstyle\int$\cr}}
}

\begin{document}

\preprint{NORDITA-2022-045}

\title{Structure constants of short operators in planar $\mathcal{N}=4$ SYM theory}
\author{Benjamin Basso}
\email{benjamin.basso@phys.ens.fr}
\affiliation{Laboratoire de Physique de l'\'Ecole Normale Sup\'erieure, ENS, Universit\'e PSL, CNRS, Sorbonne Universit\'e, Universit\'e Paris Cit\'e, F-75005 Paris, France}
\author{Alessandro Georgoudis}
\email{alessandro.georgoudis@su.se}
\affiliation{Laboratoire de Physique de l'\'Ecole Normale Sup\'erieure, ENS, Universit\'e PSL, CNRS, Sorbonne Universit\'e, Universit\'e Paris Cit\'e, F-75005 Paris, France}
\affiliation{NORDITA, Stockholm University and KTH Royal Institute of Technology, Hannes Alfv\'ens v\"ag 12, SE-106 91 Stockholm, Sweden}
\author{Arthur Klemenchuk Sueiro}
\email{arthur.klemenchuk-sueiro@phys.ens.fr}
\affiliation{Laboratoire de Physique de l'\'Ecole Normale Sup\'erieure, ENS, Universit\'e PSL, CNRS, Sorbonne Universit\'e, Universit\'e Paris Cit\'e, F-75005 Paris, France}

\date{\today}

\begin{abstract}
We present an integrability-based conjecture for the three-point functions of single-trace operators in planar $\N=4$ super-Yang-Mills theory at finite coupling, in the case where two operators are protected. Our proposal is based on the hexagon representation for structure constants of long operators, which we complete to incorporate operators of any length using data from the TBA/QSC formalism. We perform various tests of our conjecture, at weak and strong coupling, finding agreement with the gauge theory through 5 loops for the shortest three-point function and with string theory in the classical limit.
\end{abstract}

\maketitle

\section{Introduction}\label{sec:intro}

The discovery of integrability~\cite{Beisert:2010jr,Arutyunov:2009ga} in the planar limit of the $\mathcal{N}=4$ super-Yang-Mills (SYM) theory has led to tremendous advances in the study of this interacting superconformal gauge theory and of its gravitational dual, the type IIB superstring theory in $AdS_{5}\times S^{5}$~\cite{Aharony:1999ti}. The best example is the solution to the full spectrum of anomalous dimensions of single-trace operators which was argued on both sides of the duality to follow from the diagonalization of a commuting family of transfer matrices $T_{a, s}(u)$, depending on a spectral parameter $u$ and labelled by representations $(a, s)$ of the superconformal group. The latter generate an infinite number of conserved charges, pinning down the conformal primary operators, and fulfill the celebrated Hirota equation~\cite{Gromov:2009tv}
\beq\label{eq:hirota}
T_{a,s}^{+}T_{a,s}^{-}=T_{a+1,s} T_{a-1,s}+T_{a,s+1}T_{a,s-1}\, ,
\eeq
with $T_{a, s}^{\pm} = T_{a, s}(u\pm i/2)$. Once supplied with appropriate boundary conditions and analyticity requirements~\cite{Gromov:2014caa,Cavaglia:2010nm,Balog:2011nm,Gromov:2011cx}, eq.~\eqref{eq:hirota} can be cast into a set of integral Thermodynamic Bethe Ansatz (TBA) equations~\cite{Gromov:2009tv,Bombardelli:2009ns,Gromov:2009bc,Arutyunov:2009ur}, or in the more compact Quantum Spectral Curve (QSC) equations involving Q functions with simpler analytical structure~\cite{Gromov:2013pga,Gromov:2014caa}. One may expect optimistically that the remaining conformal data - the structure constants - can be addressed along the same lines and expressed in the same terms, for all values of the 't Hooft coupling, $\lambda = 16\pi^2 g^2$, as suggested by recent studies~\cite{Jiang:2019zig,Cavaglia:2021mft}.

However, to date, the known non-perturbative method for structure constants of singe-trace operators, which relies on the hexagon representation~\cite{Basso:2015zoa,Fleury:2016ykk,Eden:2016xvg,Bargheer:2017nne,Eden:2017ozn}, only works for asymptotically long operators, with very high charges. For short operators, the description is known to disagree with perturbation theory, past a certain loop order, highlighting the need to include wrapping corrections in the formalism~\cite{Basso:2017muf,Chicherin:2018avq} as in the TBA's early days~\cite{Ambjorn:2005wa,Bajnok:2008bm}. Even worse, the description is plagued with divergences that need to be subtracted carefully and it stays unknown so far how to perform this subtraction systematically, such as to make contact with the solution for the spectrum.

\begin{figure}
\begin{center}
\includegraphics[scale=0.37]{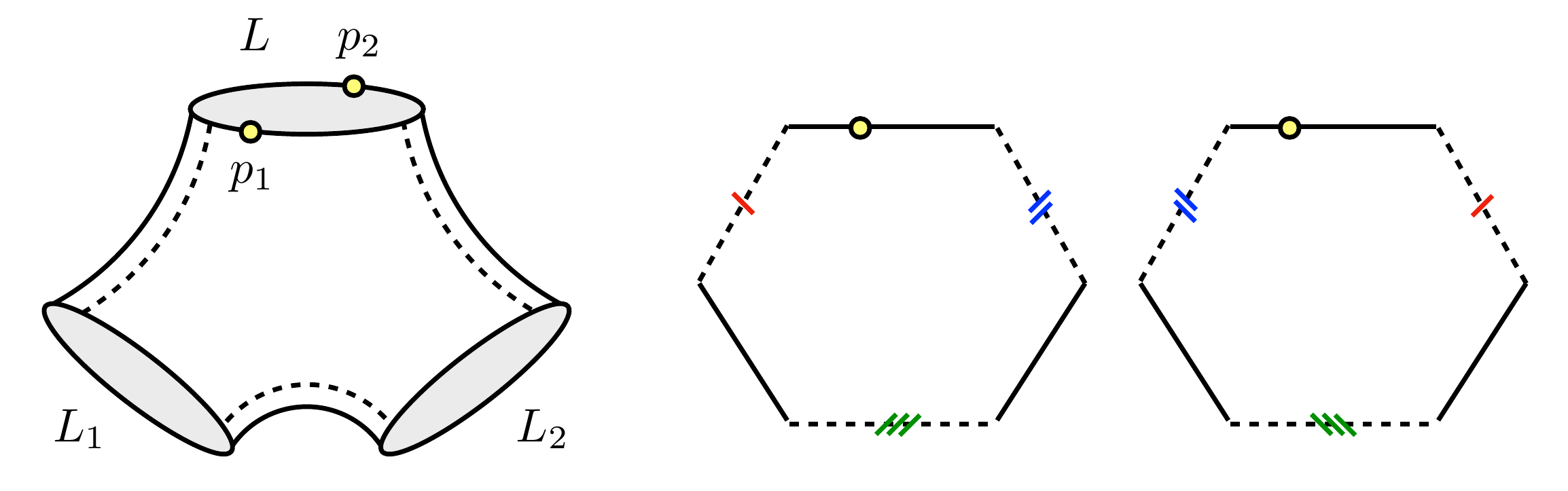}
\end{center}
%\vspace{-2cm}
\caption{Pair of pants with two empty strings at the bottom and an excited one at the top with magnons carrying momenta $p_{1}$ and $p_{2}$. The string diagram may be obtained by gluing two hexagons together along the identified mirror edges (dashed lines) as shown in the right panel.}\label{pants} 
\end{figure}

In this paper, we present a conjectured solution to this problem for a simple class of structure constants, with two half-BPS operators and one spinning operator,
\beq
C^{\circ\circ\bullet} \sim \langle \Tr\big[Z_{1}^{L_{1}}\big] \Tr\big[Z_{2}^{L_{2}}\big] \Tr\big[D^{S}Z^{L}\big] \rangle\, ,
\eeq
with $Z_{1}, Z_{2}, Z$ three complex scalar fields and $D$ a light-cone derivative. They look pictorially like in fig.~\ref{pants} with two vacua of length $L_{1}$ and $L_{2}$ merging into a length-$L$ Bethe state with $S$ magnons. Our conjecture, which extrapolates from existing results, may be viewed as a minimal way of combining the hexagon representation with the TBA/QSC spectral data, such as to obtain an exact description for operators of arbitrary length.

\section{Main conjecture}
\begin{figure}
\begin{center}
\includegraphics[scale=0.37]{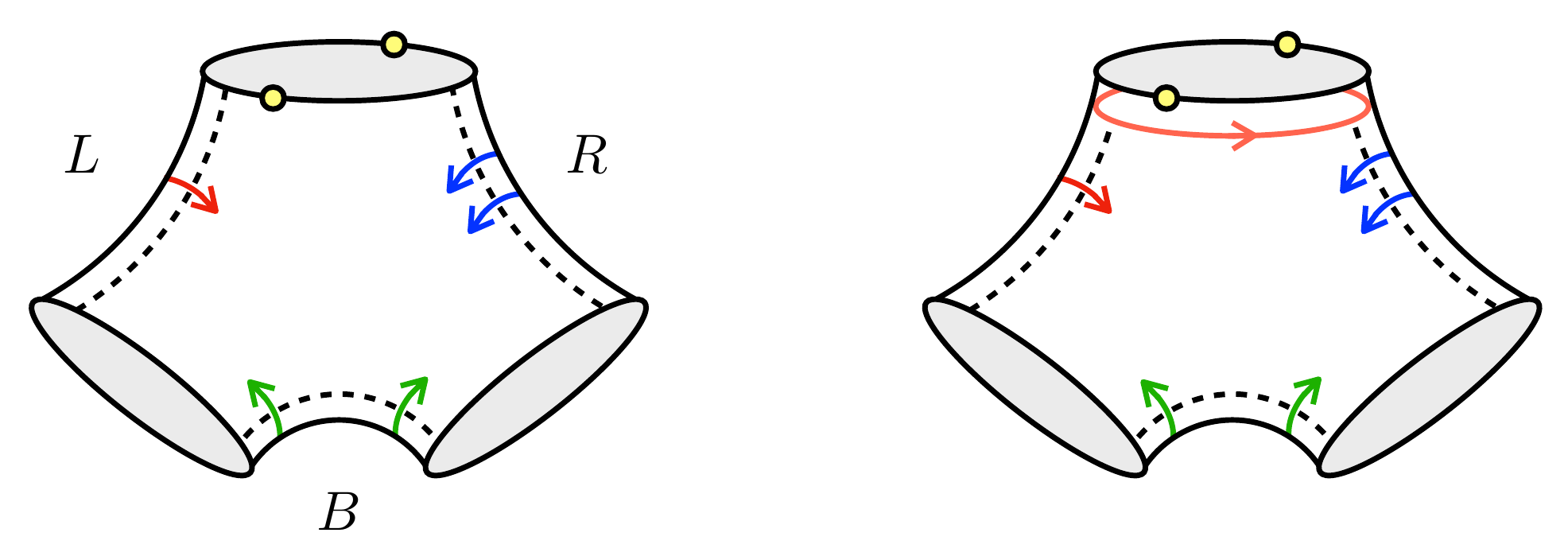}
\end{center}
%\vspace{-2cm}
\caption{Mirror magnons propagating across the left, right and bottom seams of the pair of pants. Additional magnons moving closely around the excited operator as in the right panel are needed to account for wrapping corrections.}\label{pants2} 
\end{figure}

The key idea behind the hexagon construction is that the structure constant may be obtained by attaching two hexagons together (fig.~\ref{pants}). The gluing is achieved by summing over all the states of the open strings stretching along the seams of the pair of pants. It results in a representation in terms of multiple sums of integrals, describing the multiple exchanges of particles, the mirror magnons, across the three channels of the pair of pants (fig.~\ref{pants2}), that is,
\beq\label{eq:main}
C^{\circ\circ\bullet} = \mathcal{N} \times \SumInt_{L}\times \SumInt_{R} \times \SumInt_{B} e^{-\ell_{L}\E_{L}-\ell_{R}\E_{R}-\ell_{B}\E_{B}}\,\, |\mathcal{H}|^2\, ,
\eeq
with $\N$ an overall normalization factor (see Appendix~\ref{appx:normalization}). Each process in the sums occurs with a `probability' $|\mathcal{H}|^2$, determined by the hexagon form factors~\cite{Basso:2015zoa}, and is suppressed by a factor depending on the energies of the mirror states, $\E_{L, R, B}$. The latter are conjugate to the so-called bridge lengths,
\beq
\ell_{L}= \frac{L_{1}+L-L_{2}}{2}\, , \,\, \ell_{R} = L-\ell_{L}\, , \,\, \ell_{B} = \frac{L_{1}+L_{2}-L}{2}\, ,
\eeq
fixing the distances between the hexagons.

\noindent 

The details of this construction were worked out in ref.~\cite{Basso:2015eqa} yielding definite predictions for the integrand $|\mathcal{H}|^2$ and the phase-space integral $\SumInt$. The latter reads
\beq\label{eq:complete-sum}
\SumInt = \sum_{N=0}^{\infty} \frac{1}{N!} \prod_{i=1}^{N}\sum_{a_{i} = 1}^{\infty}\int_{\C} \frac{\m_{a_{i}}(u_{i})du_{i}}{2\pi} \prod_{i<j}^{N} \p_{a_{i} a_{j}}(u_{i}, u_{j})\, ,
\eeq
with $N$ the number of magnons in the given channel and where each magnon in the sum carries a rapidity $u$ and `spin' $a$, labelling its representation under the asymptotic symmetry algebra, $\su(2|2)_{l}\oplus \su(2|2)_{r}$~\cite{Beisert:2006qh}, and its energy~\cite{Ambjorn:2005wa}
\beq\label{eq:energy}
\E_{a}(u) = \log{(x^{[+ a]}x^{[- a]})}\, ,
\eeq
with $x^{[\pm a]} = x(u\pm ia/2)$ and $x(u) = (u + \sqrt{u^2-4g^2})/2g$. The integration is done with the help of the measure
\beq\label{eq:pab}
\p_{ab}(u, v) = k_{ab}^{++} k_{ab}^{+-} k_{ab}^{-+}k_{ab}^{--}\, ,
\eeq
with $k^{\pm \pm}_{ab} = (x^{[\pm a]}-y^{[\pm b]})/(x^{[\pm a]}y^{[\pm b]}-1)$, $y^{[\pm b]} = x(v\pm i b/2)$, and with $\m_{a}$ coming from the double zero of $\p$ at $v = u$,
\beq\label{eq:zero}
\p_{ab}(u, v) \sim \m_{a}(u)^2(u-v)^2 \delta_{ab}\, ,
\eeq
with $\delta_{ab}$ the Kronecker delta. The contour of integration $\C$ runs over the real axis, up to a small detour which is explained in the next section.

The integrand $|\mathcal{H}|^2$ is less explicit, yet remarkable in that it factorizes into a number of individual weights and pairwise interactions,
\beq\label{eq:integrand}
|\mathcal{H}|^2 = \prod_{i, j, k = 1}^{N_{L, R, B}}\frac{\WL_{a_{i}}(u_{i})\WR_{b_{j}}(v_{j})\WB_{c_{k}}(w_{k})}{\p_{a_{i}b_{j}}(u_{i}, v_{j})}\, ,
\eeq
with $\{u_{i}, v_{j}, w_{k}\}$ and $\{a_{i}, b_{j}, c_{k}\}$ denoting the rapidities and spins of the magnons in the three channels. The weights $\bW_{a}(u)$ encode the interactions between the mirror magnons and the physical magnons in the excited state. As such they depend implicitly on the Bethe roots, $\bz = \{z_{1}, \ldots , z_{S}\}$, parametrizing the excited-state wave function. In the original hexagon description, see formulae in ref.~\cite{Basso:2015eqa}, they are expressed in terms of the eigenvalues of the $\su(2|2)$ transfer matrices, describing the action of the magnon S-matrices on the Bethe state~\cite{Beisert:2006qh,Arutyunov:2008zt}.

As said earlier, the problem with this description is that it is only valid in the \textit{asymptotic} limit, that is, when the length $L$ of the excited operator is large. At finite length $L$, one also expects wrapping corrections $\sim e^{-nL\E_{a}}, n=1, 2, \ldots\,$, associated with particles winding around the excited operator (fig.~\ref{pants2}). Their appearance relates to the double poles in eq.~\eqref{eq:integrand}, see eq.~\eqref{eq:zero}, for magnons with same quantum numbers, $u_{i} = v_{j}$ and $a_{i}=b_{j}$, in the left and right channels, which must be regularized somehow. An all-order derivation of these corrections, akin to the TBA for the spectrum, is still lacking. Nonetheless, one can gain several insights into the general formula from the leading exponentials, which were worked out in a number of situations~\cite{Basso:2017muf,Basso:2018cvy,Luscher}. They indicate that the hexagon formula stays intact (and as factorized as above) up to modifications of the weights $\bW$ in eq.~\eqref{eq:integrand} and normalization factor $\N$ in eq.~\eqref{eq:main}. Furthermore, the expressions hint at a simple all-order extrapolation in terms of TBA/QSC quantities.

\begin{figure}
\begin{center}
\includegraphics[scale=1.1]{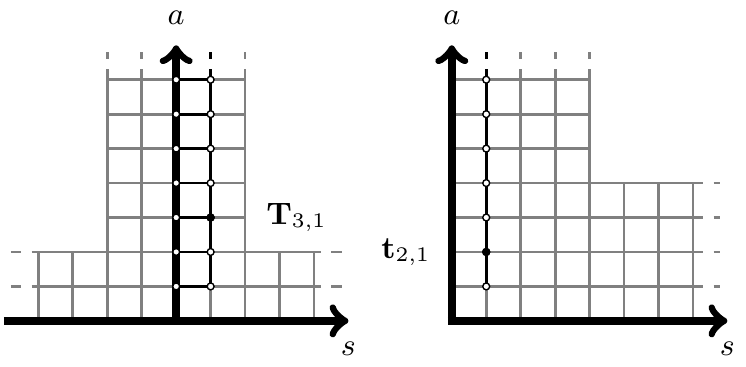}
\end{center}
\caption{Hooks for rectangular representations of $\su(2,2|4)$ and $\su(4|4)$, respectively~\cite{Kazakov:2015efa}, with highlighted the transfer matrices used in the conjecture. The $\bT/\bt$-systems are defined as solutions to the Hirota equation with null boundary conditions outside their respective domains.}\label{hooks} 
\end{figure}

To be precise, the evidence suggests that the problem is solved by 1) shifting the poles away from the real axis,
\beq\label{eq:i0}
\p_{ab}(u, v) \rightarrow \p_{ab}(u+i0, v-i0)\, ,
\eeq
in eq.~\eqref{eq:integrand},  and 2) setting
\beq\label{eq:weight-LR}
\begin{aligned}
&\mathbb{W}^{R/L}_{a}(u) = e^{\frac{1}{2}L \E_{a}(u)}\frac{\bT_{a, 1}(u)}{\bT_{a, 0}^{+/-}(u)}\, , 
\end{aligned}
\eeq
with $\bT_{a, 0}^{\pm}(u) = \bT_{a, 0}(u\pm i/2)$, and
\beq\label{eq:weight-B}
\WB_{a}(u) = e^{-\frac{1}{2}L \E_{a}(u)}\bt_{a, 1}(u)\, .
\eeq
Here, $\bt_{a,s}$ and $\bT_{a,s}$ denote the eigenvalues (for the Bethe state of interest) of two families of transfer matrices, associated with compact and non-compact representations of the superconformal algebra, respectively. The latter generate the familiar T-system underlying the TBA equations~\cite{Gromov:2009tv,Bombardelli:2009ns,Gromov:2009bc,Arutyunov:2009ur}, with $(a,s)$ labelling the nodes on the T-hook, whereas the former define an accompanying system solving the Hirota equation~\eqref{eq:hirota} on the L-hook (fig.~\ref{hooks}).

Transfer matrices in general are only defined up to gauge transformations~\cite{Gromov:2009tv}
\beq
T_{a, s} \rightarrow g_{1}^{[a+s]} g_{2}^{[a-s]} g_{3}^{[s-a]} g_{4}^{[-a-s]} \, T_{a, s}\, ,
\eeq
which leave eq.~\eqref{eq:hirota} invariant $\forall g_{j}$, with $g_{j}^{[n]} = g_{j}(u+in/2)$. In our case, the gauge is fixed, with $\bT$ given in the distinguished $\bT$-gauge of refs.~\cite{Gromov:2014caa,Gromov:2011cx} and with $\bt$ normalized as in ref.~\cite{Kazakov:2015efa}, up to a sign, $\bt_{a,s|\textrm{here}} = (-1)^{as} \bt_{a,s|\textrm{there}}$. One may also find in these references explicit representations for the transfer matrices in terms of the  QSC's Q functions, which are valid for operators of any length $L$ and prove extremely useful in practical applications.

Alternatively, one may state the conjecture in a gauge-invariant way, using the Y functions,
\beq\label{eq:prod-W}
Y_{a, s} = \bT_{a, s+1}\bT_{a, s-1}/\bT_{a+1, s}\bT_{a-1, s}\, .
\eeq
This is so at least for $\WR_{a}$ and $\WL_{a}$, which readily obey
\beq\label{eq:ratio-W}
e^{-L\E_{a}}\WR_{a} \WL_{a} = \frac{\bT_{a, 1}\bT_{a,1}}{\bT^{+}_{a, 0}\bT^{-}_{a,0}} = \frac{Y_{a, 0}}{1+Y_{a, 0}}\, ,
\eeq
using eq.~\eqref{eq:hirota} and the left-right symmetry, $\bT_{a,s} = \bT_{a, -s}$, observed for our states. Integral representations~\cite{Kazakov:2015efa} yield the extra relation
\beq
\log{\left[\frac{\WL_{a}(u)}{\WR_{a}(u)}\right]} = i\sum_{b=1}^{\infty}\dashint_{\C} \frac{dv}{2\pi} L_{b}(v) \partial_{v} \log{\p_{ba}(v, u)}\, ,
\eeq
with $L_{a}(u) = \log{(1+Y_{a, 0}(u))}$ and with a principal-value prescription for the pole at $v=u$ when $b=a$. The analysis is much harder for the $\bt$'s however, which are less obviously embedded in the TBA formalism.

Formulae~\eqref{eq:weight-LR} and~\eqref{eq:weight-B} are our main results for the resummation of the wrapping corrections. We stress that they go along with the prescription~\eqref{eq:i0}, with the $\pm i0$ shifts responding to the $\mp$ shifts of $\bT_{a, 0}$ in eq.~\eqref{eq:weight-LR}.

At last, there is a formula for the normalization constant, $\N$, which depends on the Bethe state alone. It is conjectured to be given, up to a simple factor, by Fredholm determinants canonically associated with the TBA equations, in line with recent findings for structure constants of determinant operators~\cite{Jiang:2019xdz,Jiang:2019zig}. Its detailed presentation is deferred to the Appendix~\ref{appx:normalization}.

\begin{figure}
\begin{center}
\includegraphics[scale=0.4]{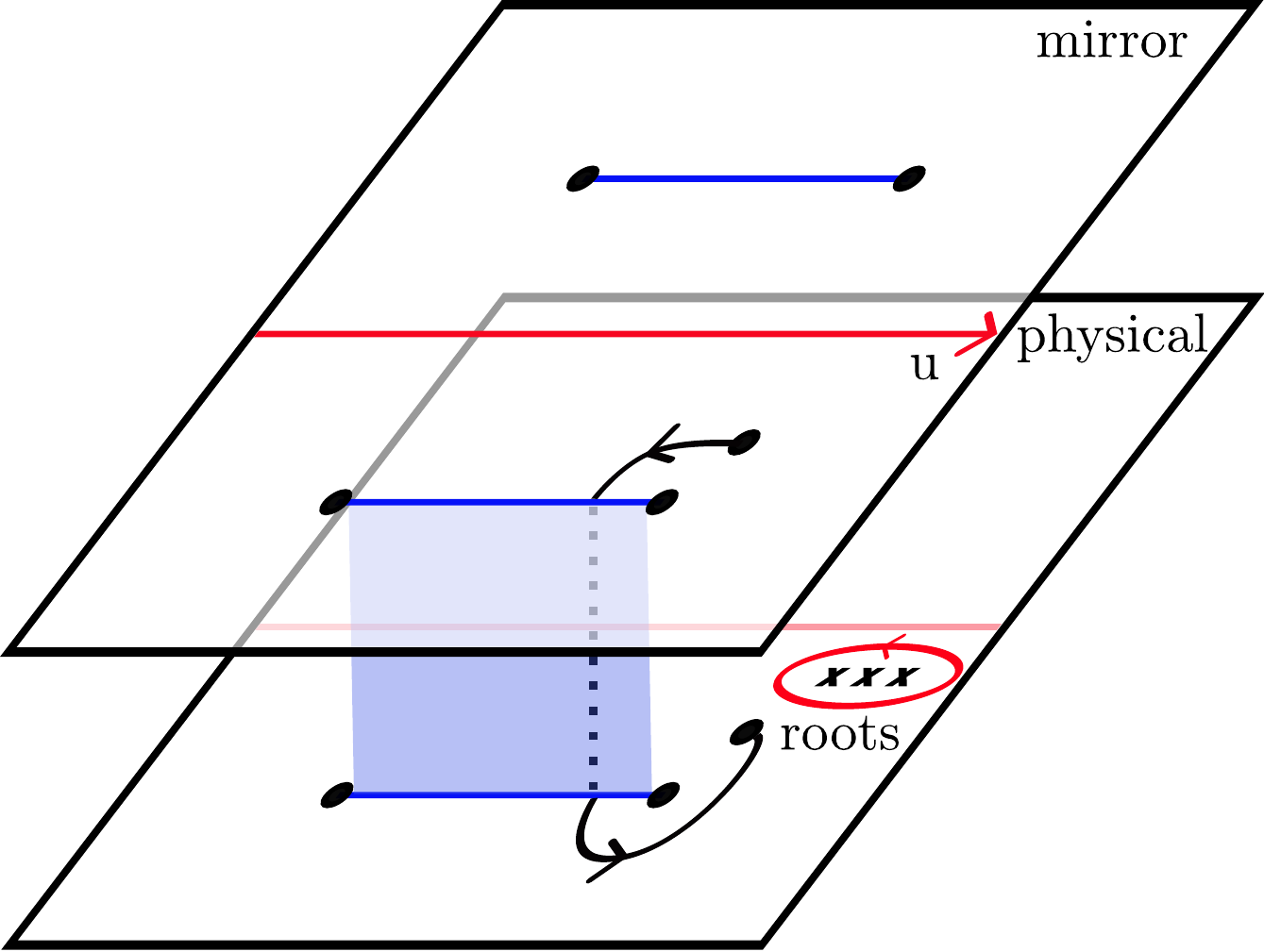}
\end{center}
%\vspace{-2cm}
\caption{Rapidity sheets of $\WR_{1}(u)$ with branch points at $u = \pm 2g \pm i/2$ and with poles on the second sheet at the positions of the Bethe roots. The contour $\C$ must go along the real line in the mirror sheet and around the roots in the physical sheet.}\label{contour} 
\end{figure}

\section{Contour and partitions}

In addition to the mirror sums, the hexagon rules also predict a sum over all possible ways of distributing the physical magnons on the two sides of the pair of pants.
This sum was skipped in the formula above, which may be viewed as describing a configuration with all roots standing on the same hexagon. The reason for this omission is that this sum is not independent and may be restored for a suitable choice of the contour $\C$ in eq.~\eqref{eq:complete-sum}, as was first noticed in refs.~\cite{Jiang:2015lda,Jiang:2016ulr}. The key observation is that the fundamental weights (with $a=1$) develop simple poles on the physical sheet, at the locations of the Bethe roots, $\log{\bW_{1}^{\textrm{phys}}(u)}\sim \log{(u-z_{i})}$ for $u\sim z_{i}$. In analogy with the analytic continuation trick used in the Thermodynamic Bethe Ansatz~\cite{Dorey:1996re,Bazhanov:1996aq,Fioravanti:1996rz}, see also refs.~\cite{Pozsgay:2014gza,Jiang:2019xdz,Jiang:2019zig} for recent discussions, the contour must be chosen to go around these Bethe poles.

There are two alternatives: $\C$ may go around the poles of $\WR_{1}$ sitting on the second sheet of $x(u+i/2)$, as depicted in figure~\ref{contour}, or around those of $\WL_{1}$ on the second sheet of $x(u-i/2)$. They may be shown to be equivalent, for roots solving the exact Bethe equations, $Y_{1, 0}^{\textrm{phys}} = -1$. Either way, extracting the residues has the desired effect of moving magnons to the next hexagon, over the right or left bridge.

In particular, keeping \textit{only} the integrals around the roots one reproduces the sum over the partitions~\cite{Basso:2015zoa,Escobedo:2010xs}
\beq
C^{\circ\circ\bullet} \rightarrow \N \times \sum_{\alpha} \frac{(-1)^{|\alpha|} \prod_{z_{j}\in \alpha}e^{ip(z_{j})\ell_{R}}\mathcal{T}(z_{j})}{\prod_{z_{i}, z_{j}\in \bar{\alpha}, \alpha}h(z_{i}, z_{j})}\, ,
\eeq
which runs over the subsets of $\bz$, with $\alpha \cup \bar{\alpha} = \bz$, $p(u) =  i\E_{1}^{\textrm{phys}}$ the spin-chain momentum and $h(z_{i}, z_{j})$ the physical hexagon form factor~\cite{Basso:2015zoa}. $\mathcal{T}(z_{j})$ is a dressing factor collecting the wrapping corrections to the residue of $\WR_{1}(u)^{\textrm{phys}}$ at $u = z_j$. Using eqs.~\eqref{eq:prod-W} and~\eqref{eq:ratio-W}, it may be written as
\beq
\mathcal{T} = \exp\, \bigg[\frac{i}{2}\Phi(z_{j})-\frac{i}{2}\sum_{a=1}^{\infty} \int \frac{du}{2\pi} L_{a}(u) \partial_{u}\log{p_{a1}(u^{\gamma}, z_{j})}\bigg]\, ,
\eeq
where $\Phi$ is the wrapping-induced phase shift in the exact Bethe equations~\cite{Bajnok:2008bm} and $p_{a1}(u^{\gamma}, z_{j})$ is the measure~\eqref{eq:pab} in the mixed, mirror-physical kinematics, see ref.~\cite{Basso:2015zoa} for notations. It agrees perfectly with the L\"uscher formula proposed in ref.~\cite{Basso:2017muf}, in the IR limit $L_{a} \approx Y_{a, 0} \sim e^{-L\E_{a}}$, up to an overall factor which follows from $\N$.

At last, let us comment on $\WB_{1}$. In the asymptotic description, this weight also appears to have poles on the physical sheets. It suggests the strange possibility that the physical magnons should also be transported over the bottom bridge, in tension with the locality of the hexagon rules. However, the tension goes away when wrapping corrections are included. The exact expression~\eqref{eq:weight-B} is indeed free from singularities, since transfer matrices are regular everywhere on their Riemann surfaces~\cite{Gromov:2014caa,Kazakov:2015efa}. The puzzle arises because the analytic continuation to the physical sheets does not commute with the large $L$ limit; the mirror asymptotic formula being valid only on the mirror sheet. The effect is nicely illustrated with the fundamental transfer matrix at weak coupling. On the physical sheet, it is given as a sum of two terms,
\beq
\bt_{1,1}^{\textrm{phys}} \propto e^{\frac{i}{2}pL}(\frac{Q(u+i)}{Q(u)} -1 ) + e^{-\frac{i}{2}pL} (\frac{Q(u-i)}{Q(u)} -1 )\, ,
\eeq
with $Q = \prod_{j=1}^{S}(u-z_{j})$ and $e^{ip} = (u+i/2)/(u-i/2)$, and it is regular at $u= z_j$, despite the poles in each term, in virtue of the Baxter equation. Nonetheless, the cancellation cannot be seen asymptotically from the mirror kinematics, because the two terms map then to different orders $e^{\pm ipL/2} \rightarrow e^{\mp \E L/2}$ at large $L$.

\section{Advanced checks}

We turn to direct tests of our conjecture with available data from field and string theory. To simplify the discussion, we shall focus on a ratio of structure constants
\beq\label{eq:ratio}
R (\ell_{B}) = C^{\circ\circ\bullet} / \lim_{\ell_{B} \rightarrow \infty} C^{\circ\circ\bullet}\, ,
\eeq
with the limit taken at fixed $L$. Owing to the factorization in~eqs.~\eqref{eq:main} and~\eqref{eq:integrand}, this ratio receives no contributions from the left-right channels or normalization factor. As such, it is given by a single sum of mirror integrals,
\beq\label{eq:ratio2}
R = 1 + \sum_{a=1}^{\infty} \int \frac{du}{2\pi}\m_{a}(u) e^{-\ell_{B}\E_{a}(u)} \WB_{a}(u) + \ldots\, ,
\eeq
with dots standing for terms with $N=2, 3, \ldots$ magnons.

\textit{Weak coupling.} Taking the weak coupling limit, $g^{2}\rightarrow 0$, greatly simplifies the analysis, by suppressing the $N$-particle terms in eq.~\eqref{eq:ratio2} which kick in at $N(\ell_{B}+N)$ loops at the earliest~\cite{Basso:2015eqa}. The first wrapping correction, on the other hand, enters at $(L+\ell_{B}+2)$ loops and thus dominates over the higher-particle terms for $L < \ell_{B}+2$. This hierarchy is observed for the `shortest' three-point function of two stress tensors and one Konishi operator, with $\ell_{B}=1$ and $L = S = 2$. The associated ratio \eqref{eq:ratio} was studied in refs.~\cite{Georgoudis:2017meq,Chicherin:2018avq} through higher loops, using diagrammatic techniques, and a mismatch was found with the original hexagon formula at five loops, that is, precisely when the first wrapping occurs. The discrepancy reads~\cite{Chicherin:2018avq} 
\beq\label{eq:5loops}
\frac{\delta R}{g^{10}} = 972 \zeta_3 - 2700 \zeta_5 + 5355 \zeta_7 - 2376 \zeta_3 \zeta_5 - 1512 \zeta_9\, ,
\eeq
with $\zeta_{n} = \sum_{a=1}^{\infty}a^{-n}$ the Riemann zeta function. Nicely, formula~\eqref{eq:weight-B} restores the agreement with the field theory. To see that, one needs the expression for $\bt_{a, 1}$. It is easily obtained using the QSC~\cite{Kazakov:2015efa}
\beq
\bt_{a, 1}(u) = -\sum_{j=1}^{4} \bQ_{j}(u+ia/2) \tilde{\bQ}^{j}(u-ia/2)\, ,
\eeq 
with $\bQ_{j}, \tilde{\bQ}^{j}$ the fundamental (fermionic) Q functions. The weak coupling expansion may then be done efficiently using powerful QSC solvers~\cite{Marboe:2014gma,Gromov:2015vua,Marboe:2018ugv}. The outcome is a complicated meromorphic function of the rapidity, which is too lengthy to be shown here. Its integration is immediate though and reproduces perfectly the gauge-theory prediction, including the wrapping contribution~\eqref{eq:5loops}.

\textit{Strong coupling.} The ratio~\eqref{eq:ratio} may also be calculated in the string theory at strong coupling, $g\gg 1$, in the classical limit $L, S, \ell_{B} \sim g$. In this regime, one may rely on the integrability of the classical worldsheet theory to get a closed-form expression for the structure constants~\cite{Kazama:2016cfl,Kazama:2011cp}. It is given in terms of the eigenvalues of the string monodromy matrix $\Omega(x) \in SU(4|4)$, generating the conserved charges of the string~\cite{Kazakov:2004qf,Beisert:2005bm,Bena:2003wd},
\beq
\Omega \cong \textrm{diag} (e^{i\pS_{1}}, e^{i\pS_{2}},e^{i\pS_{3}}, e^{i\pS_{4}}|e^{i\pA_{1}}, e^{i\pA_{2}}, e^{i\pA_{3}}, e^{i\pA_{4}})\, ,
\eeq
with $x$ the spectral parameter and $\pS_{i}(x), \pA_{j}(x)$ the so-called quasi-momenta. General expressions for the states of interest give~\cite{Gromov:2009tq,Kazakov:2004nh}
\beq\label{eq:sl2}
\pA_{1}(x) = -\pA_{4}(x) = -\pA_{2}(\tfrac{1}{x}) = \pA_{3}(\tfrac{1}{x}) = \frac{L}{2i}\E(x) + G(x)\, ,
\eeq
and similarly for $\pS_{i}$ with $G\rightarrow 0$, with $\E(x) = i x/(g(x^2-1))$ and with $G(x)$ the resolvent of the classical curve. The string theory then predicts that~\cite{Kazama:2016cfl,Kazama:2011cp}
\beq\label{eq:string}
\log R_{\textrm{string}}  = \int_{U^{+}}\frac{du(x)}{2\pi} \sum_{j=1}^{4} \big[\textrm{Li}_{2}(\xi e^{i\pA_{j}})- \textrm{Li}_{2}(\xi e^{i\pS_{j}})\big]\, ,
\eeq
at large $g$, with $\textrm{Li}_{2}(z) = \sum_{n=1}^{\infty} z^{n}/n^2$ the dilogarithm, $U^{+}$ the upper half of the circle $|x| = 1$, $u(x) = g(x+1/x)$, and $\xi(x) = e^{-\frac{1}{2}(L_{1}+L_{2}) \E(x)}$.

Now, the hexagon sum in eq.~\eqref{eq:ratio2} also greatly simplifies in this regime. It exponentiates and its exponent can be calculated exactly using the clustering method~\cite{Jiang:2016ulr}
\beq\label{eq:lnR-strong}
\log R_{\textrm{strong}} = \int_{U^{+}}\frac{du(x)}{2\pi} \int_{0}^{\xi} \frac{dq}{q} \log{\bigg[\sum_{a=0}^{\infty}q^{a}\bt_{a, 1}(x)\bigg]}\, ,
\eeq
with $\bt_{0, 1} = 1$. In this form, the comparison with the string formula is immediate. One first observes~\cite{Gromov:2010vb,Gromov:2010km} that in the classical limit the $\bt$'s become characters of the group element $\Omega$ in finite-dimensional representations of $\su(4|4)$, with the generating function
\beq\label{eq:generator}
\sum_{a=0}^{\infty}q^{a}\, \bt_{a, 1} = \textrm{sdet} (1-q\, \Omega(x)) = \prod_{j=1}^{4} \frac{1-q\, e^{i\pS_{j}(x)}}{1-q\,  e^{i\pA_{j}(x)}}\, ,
\eeq
and `sdet' the Berezinian (superdeterminant). Plugging then this expression inside eq.~\eqref{eq:lnR-strong} and integrating over $q$, one gets
\beq
\log{R_{\textrm{strong}}} = - \int_{U^{+}}\frac{du(x)}{2\pi} \, \textrm{str}\,\big[ \textrm{Li}_{2}(\xi \, \Omega)\big]\, ,
\eeq
with `str' the graded trace on the $\textbf{4}|\textbf{4}$ module, in perfect agreement with the string result~\eqref{eq:string}.

It is also instructive to look at these expressions in the asymptotic limit, which corresponds here to $L\gg g$. In this regime, half of the eigenvalues are exponentially large, see eq.~\eqref{eq:sl2},
\beq
e^{i\pA_{1, 2}} \sim e^{i\pS_{1, 2}}  \sim e^{L\E/2} \, ,
\eeq
implying that the product in eq.~\eqref{eq:generator} and the sum in eq.~\eqref{eq:string} can be restricted to $j=1,2$, as for the $\su(2|2)$ problem studied in ref.~\cite{Jiang:2016ulr}. The four remaining eigenvalues, with indices $j =3,4,$ are exponentially small $\sim e^{-L\E/2}$ and stand for wrapping corrections. This splitting makes it clear that maintaining the agreement with the classical string theory away from the large $L$ limit amounts to restoring the full $\su(4|4)$ symmetry, as done in eq.~\eqref{eq:weight-B} through the replacement of the `old' $\su(2|2)$ hexagon weights by `new' $\su(4|4)$ ones.

Finally, let us mention that similar tests may be performed for the left-right sums and normalization factor. The analysis is significantly harder and not as transparent as the one carried out here, but the comparison may be done order by order in the wrapping parameter $\sim e^{-L\E}$, using the character solution for the $\bT$-system~\cite{Gromov:2009tq,Gromov:2010vb,Gromov:2010km}.

\section{Conclusion and outlook}

We reported a conjecture for structure constants of single-trace operators in planar $\N=4$ SYM theory at finite coupling and found evidence for it at weak and strong coupling from matching with gauge and string theory. 

For simplicity we focused in this paper on operators in the smallest closed subsector. It seems possible however to extend the conjecture to the higher-rank sectors, such as to describe operators with derivatives but also scalars and fermions. Structure constants of this sort were analyzed in ref.~\cite{Basso:2017khq} in the asymptotic limit and found to be subject to a flavor selection rule, imposing a stringent left-right symmetry on the operators, that is, $\bT_{a, s} = \bT_{a, -s}$~\cite{Gromov:2014caa}. We believe our formula applies as well to this more general case. It is less clear how to extend it to asymmetric operators though, should the left-right symmetry be broken by wrapping corrections. It would also be fascinating to explore applications of our findings to structure constants with two or more unprotected operators, like the ones studied recently in refs.~\cite{Bianchi:2019jpy,Bercini:2020msp,Bercini:2021jti,Bianchi:2022oyz,Caetano:2016keh}.

One may also expect applications in lower dimensional AdS/CFT setups, where the hexagon formalism was developed~\cite{Eden:2021xhe}, or in integrable deformations of the SYM theory, such as the fishnet theory~\cite{Gurdogan:2015csr,Caetano:2016ydc}. The latter is of particular interest to make contact with approaches based on the Separation of Variables~\cite{Cavaglia:2021mft,Derkachov:2019tzo} or to shed light on the group theory meaning of our formula.

Admittedly, the evaluation of the hexagon sums at high orders in perturbation theory or at finite coupling necessitates a fair amount of work. One will also run into serious difficulties at special points in the parameter space, like at the extremal points where the mixing with double-trace operators is important. At these points the hexagon integrals no longer make sense and must be analytically continued. The Pfaffian structure identified in ref.~\cite{Basso:2017khq} may help simplifying the algebra. It may also help obtaining a more compact representation for the structure constants, as found recently in the study of large-charge four-point functions~\cite{Kostov:2019stn,Belitsky:2020qrm,Coronado:2018ypq}.

%%%%%%%%%%% Acknowledgements

\begin{acknowledgments}
We are indebted to D.~Volin for several nice discussions and for an illuminating introduction to the Quantum Spectral Curve. We thank S.~Ekhammar, V.~Gon\c calves, A.~K.~Kashani-Poor, V.~Kazakov, R.~Klabbers, M.~Preti, I.~Sz\'ecs\'enyi, P.~Vieira for interesting discussions. We are very grateful to the Galileo Galilei Institute and to the scientific program \textit{Randomness, Integrability, and Universality} for support and hospitality during the final stage of this project. This work received support from the French National Agency for Research grant ANR-17-CE31-0001-02. Nordita is supported in part by NordForsk.
\end{acknowledgments}

\appendix

\section{Normalization factor}\label{appx:normalization}

In this appendix, we present the conjecture for the normalization factor $\N$ in eq.~\eqref{eq:main}. In the large $L$ limit, this factor is given by the ratio~\cite{Basso:2015zoa}
\beq\label{eq:N-asy}
\N\,^2 \approx \frac{\prod_{i=1}^{S}\mu(z_{i})\prod_{i < j}^{S}p(z_{i}, z_{j})}{G}\, ,
\eeq
where the numerator is the physical analog of the measure in eq.~\eqref{eq:complete-sum} evaluated on the roots $\bz = \{z_{1}, \ldots , z_{S}\}$. The denominator is the Gaudin determinant derived from the asymptotic Bethe ansatz equations~\cite{Beisert:2005fw}, see ref.~\cite{Basso:2015zoa} for the explicit expression. At finite $L$, we expect it to be replaced by suitable Fredholm determinants associated with the TBA equations. The expectation is based on known results for boundary entropies, or $g$-functions \cite{Dorey:2004xk,Pozsgay:2010tv,Jiang:2019xdz,Kostov:2018dmi,Caetano:2021dbh}, which display the same determinants and were shown to relate to structure constants in refs.~\cite{Jiang:2019xdz,Jiang:2019zig}. Fredholm determinants also appear naturally in the hexagon setup, like for diagonal structure constants in the planar fishnet theory \cite{Gurdogan:2015csr}, which were analyzed in ref.~\cite{Basso:2018cvy} with the help of the Leclair-Mussardo formula~\cite{Leclair:1999ys,Pozsgay:2010xd}. At last, the L\"uscher formula of ref.~\cite{Basso:2017muf} hints at their emergence in our problem, in the IR limit $Y_{a, 0}\sim e^{-L\E_{a}}$.

To introduce these determinants, recall first that the TBA equations can be cast into the form
\beq\label{eq:TBA}
\log{\Y_{a, s}(u)} = -L\E_{a}(u) \delta_{s, 0}+ \sum_{b, t} \int_{\C} \frac{dv}{2\pi} \L_{b, t}(v) K^{b, t}_{a, s}(v, u)\, ,
\eeq
for some kernels $K$ and with $\L_{a, s} = \log{(1+\Y_{a, s})}$, where $\Y_{a, s} = Y_{a, s}$ for $|s| <2$ and $\Y_{a, s} = 1/Y_{a, s}$ otherwise, and with $Y_{a,s}$ the Y functions introduced earlier. The contours here are those suitable for the excited state under consideration~\cite{Gromov:2009bc,Bombardelli:2009ns,Arutyunov:2009ur}. In particular, for $(a, s) = (1, 0)$, $\C$ includes a loop around the Bethe roots, solving $1+\Y^{\textrm{phys}}_{1,0} = 0$.

One may then define the Fredholm determinant $\D$ for the full set of TBA equations, using
\beq\label{eq:logD}
\log{\D} = -\sum_{N=1}^{\infty} \frac{1}{N} \int_{\C} \frac{du_{1}\ldots du_{N}}{(2\pi)^{N}} \textrm{Tr} \bigg[ \overrightarrow{\prod_{i}}\bold{K}(u_{i}, u_{i+1})\bigg] \, ,
\eeq
where $u_{N+1} = u_{1}$ and with $\bold{K}(u, v)$ an infinite matrix with elements
\beq
\bold{K}_{a, s}^{b, t}(u, v) = 2\pi \frac{\delta \log{\Y_{a,s}(u)}}{\delta \log{\Y_{b, t}(v)}} =  \frac{\Y_{b, t}(v)}{1+\Y_{b, t}(v)} K_{a, s}^{b, t}(v, u)\, .
\eeq
The trace in eq.~\eqref{eq:logD} runs over all the labels $(a, s)$ in the Y system. We define in a similar manner the right/left minor $\D_{r/l}$ by restricting the second index `$s$' in the trace to positive/negative values.

We may now state the finite-$L$ conjecture for $\N$,
\beq\label{eq:N-appx}
\begin{aligned}
&\N\,^{2} = \frac{\D_{l}\D_{r}}{\D}  \times\\
& \exp{-\frac{1}{2}\sum\limits_{a,b=1}^{\infty} \dashint_{\C} \frac{du dv}{(2\pi)^2}L_{a}(u) L_{b}(v) \partial_{u}\partial_{v}\log{\p_{ab}}(u, v)}\, ,
\end{aligned}
\eeq
where in the second line $L_{a} = \log{(1+Y_{a, 0})}$ and with a principal-value integration for the double pole at $v=u$ when $b=a$. The ratio of determinants is used here to mod out an unwanted `entropy' coming from the auxiliary Y functions, $\Y_{a, s \neq 0} = \Op(1)$ at large $L$ or $u$~\cite{Gromov:2009tv}. This choice also appears natural in light of the results found asymptotically for higher-rank structure constants~\cite{Basso:2017khq}, with $\D$-factors mapping to Gaudin (sub-)determinants of the nested Bethe ansatz equations. The Gaussian factor in the second line of eq.~\eqref{eq:N-appx} should on the other hand be viewed as defining the finite-length counterpart of the numerator in eq.~\eqref{eq:N-asy}.

As a sanity check, one may verify that eq.~\eqref{eq:N-appx} reduces to eq.~\eqref{eq:N-asy} when $L\gg 1$. The relevant contributions come from the contour integrals around the roots. The analysis for the $\D$'s can be derived from refs.~\cite{Jiang:2019xdz,Jiang:2019zig}, giving
\beq
{\rm Fredholm} \rightarrow \frac{\prod_{S=1}^{M} \rho(z_{j})}{G}\, ,
\eeq
with $\rho(u) = -i\partial_{u}\ln{\Y^{\textrm{phys}}_{1}}$. The Gaussian factor is no more difficult to treat, though some care is needed to deal with the double pole in the integrand. It yields
\beq
{\rm Gauss} \rightarrow \prod_{i=1}^{S} \frac{\mu(z_{i})}{\rho(z_{i})} \times \prod_{i < j}^{S}p(z_{i}, z_{j}) \, ,
\eeq
up to an irrelevant phase. One immediately observes that the $\rho\,$-terms cancel out in the product, leaving us with eq.~\eqref{eq:N-asy}. One may also work out the leading IR corrections $\sim e^{-L\E}$ and verify after some algebra that it agrees with the L\"uscher formula in ref.~\cite{Basso:2017muf}.

%%%%%% Bibliography

\bibliography{biblio_c123}
\end{document}